\documentclass[12pt]{iopart}

\usepackage{iopams}
\usepackage{graphicx}
\begin{document}

\title[Speed controlling and filtering of optical pulses in Bose condensates]
        {Magnetic field dependence and the possibility of filtering ultraslow light pulses in atomic gases with Bose–Einstein condensates}

\author{Andrii Sotnikov}

\address{A.I. Akhiezer Institute for Theoretical Physics, NSC KIPT,
1 Akademichna Street, Kharkiv 61108, Ukraine}
\ead{a\_sotnikov@kipt.kharkov.ua}
\begin{abstract}
This paper studies theoretically the ultraslow light phenomenon in Bose–Einstein condensates of alkali-metal atoms. The description is based on the linear approach that is developed in the framework of the Green function formalism. It is pointed out that the group velocity of light pulses that are tuned up close to resonant lines of the alkali atoms' spectrum may strongly depend on the intensity of the external static magnetic field. The possibility of filtering optical pulses using the ultraslow light phenomenon in Bose condensates is discussed.
\end{abstract}

%Uncomment for PACS numbers title message
\pacs{03.75.Hh, 32.60.+i, 42.25.Bs}
% Keywords required only for MST, PB, PMB, PM, JOA, JOB?
%\vspace{2pc}
%\noindent{\it Keywords}: Article preparation, IOP journals
% Uncomment for Submitted to journal title message
\submitto{\PS}
% Comment out if separate title page not required
%\maketitle

\section{Introduction}
Bose-Einstein condensate (BEC) is one of the most impressive
examples when the matter demonstrates its quantum nature on the
macroscopic level. Now this system is interesting also due to the
possibility to observe the electromagnetic pulses propagating with
extremely slow group velocities in it~\cite{Hau1999N}. Evidently, the
choice of alkali-metal vapors corresponds to the fact that a behavior of a
many-particle system consisting of alkali metals is more predictable
and convenient from the standpoint of quantum mechanics than a
behavior of the system consisting of atoms with more complicated
structure. The main reason for that is a hydrogenlike structure of
alkali metals. Therefore, the energy spectrum of these atoms and
interaction peculiarities with an external electromagnetic field are
well-studied to the present moment.

For the many-particle system consisting of alkali atoms one can
construct rather exact microscopic description, which includes the
effects of collective interaction of these systems with
electromagnetic waves. This description is based on an approximate
formulation of the second quantization method~\cite{Pel2005JMP} in
the case of a presence of bound states of particles. In the
framework of this formulation one may consider alkali metals as
bound states of two ''elementary'' particles: the valence electron and
the atomic core. This method allows not to consider an atom as a
compound particle, but describe it as an elementary object with
preserving the information relating to its internal degrees of freedom
(wave functions, energy spectrum, etc).

On the basis of the developed microscopic approach in
Refs.~\cite{Sly2006CMP,Sly2008PRA,Sly2009PLA} it was studied the response
of a system to an external electromagnetic perturbation in the
framework of the Green-function formalism. In terms of the Green
functions the expressions for the permittivity and magnetic
permeability of an ideal gas of hydrogenlike atoms with a
Bose condensate. It was found that the permittivity demonstrates the
resonance behavior at frequencies close to the allowed
transitions between certain quantum states of an atom. Note that
this behavior of the permittivity is well-known from the
classical approaches~\cite{Scully2002}. But, the proposed theory allows
to estimate the values of all microscopic quantities that appear
in the expressions characterizing the response of a gas with a BEC.

It it easy to see that the energy spacing between different states
of an atom can strongly depend on the intensity of an external
magnetic field. Aside the cases, in which this field can be
specially turned on, it can appear due to the usage of
magneto-optical traps and bias fields in BEC-related experiments.
The mentioned dependence of the energy spectrum on the magnetic
field corresponds to the Zeeman splitting (or the Paschen-Back
splitting) of the ground and excited states of an atom. In turn, the
resonance behavior of the refractive index is one of the necessary
conditions for the pulses slowing in a Bose
condensate~\cite{Sly2008PRA}. Thus, one can expect that by the
change of the intensity of a bias field it is possible to control
the propagation velocity and to filter light pulses that propagate
through a BEC cloud.

\section{Linear response of a gas to an external probe radiation}
Firstly, it should be noted that in
the present paper the description is based on the
weak-field approximation. In other words,
it is assumed that an external electromagnetic field (laser pulse)
is probe and practically do not influence on
the occupation of the states in the system. This limitation is
necessary for the usage of the results of the linear-response
theory, which is based on the Green-functions formalism
\cite{Akhiezer1981}. The mentioned fact excludes the possibility to
describe the influence of an additional coupling laser (quantum
interference effects) on the system
under consideration. Note that the coupling field was used in the
experiment~\cite{Hau1999N} to provide an electromagnetically-induced
transparency (EIT) at the resonance regions. But, as it is shown in
Ref.~\cite{Sly2008PRA}, the usage of quantum interference effects
is not a necessary condition for the pulses slowing.
Moreover, the coupling field may influence on the energy spectrum
of the atoms. Thus, it inevitably produce additional difficulties
in the construction of the description.

Usually, for the achievement of the BEC regime in the systems
consisting of alkali-metal atoms, the dilute vapors with the density
of particles $\nu\lesssim10^{14}$~cm$^{-3}$ are used. These vapors
with a good precision can be considered as ideal gases
\cite{Pitaevskii2003}. Hence, to describe microscopically the
propagation of electromagnetic pulses in this medium, it is
convenient to use the results \cite{Sly2008PRA}. There it was shown
that the expression for the permittivity of the gas in the BEC state
can be written as follows:
\begin{equation}                                                            %%%% ---- %%%% Eq. 1
    \epsilon^{-1}(\textbf{k},\omega)
    \approx 1+\frac{4\pi}
    {k^2}\sum\limits_{\alpha,\beta}
    \frac{(\nu_{\alpha}-\nu_{\beta})
    |\sigma_{\alpha\beta}(\textbf{k})|^2}
    {\omega+\Delta\varepsilon_{\alpha\beta}
    +i\gamma_{\alpha\beta}},\label{eq.m-2.1}
\end{equation}
where $\omega$, $\textbf{k}$ are the frequency and wave vector of
the external field (probe laser), $\nu_{\alpha}$ is the density
of condensed atoms in the $\alpha$ quantum state,
$\Delta\varepsilon_{\alpha\beta}$ is the energy interval between the
$\alpha$ and $\beta$ states, and $\gamma_{\alpha\beta}$ is the linewidth
related to the probability of a spontaneous transition between these
states. The quantity $\sigma_{\alpha\beta}(\textbf{k})$ is the matrix
element of the charge density, which is defined by the relation
\begin{equation}\label{eq.m-2.2}                                           %%%% ---- %%%% Eq. 2
    \sigma_{\alpha\beta}(\textbf{k})
    =e\int d\textbf{y}\varphi_{\alpha}^{*}(\textbf{y})
    \varphi_{\beta}(\textbf{y})
    \left[\exp{\left(i\frac{m_{p}}{m}
    \textbf{k}\textbf{y}\right)}
    -\exp{\left(-i\frac{m_{e}}{m}
    \textbf{k}\textbf{y}\right)}\right],
\end{equation}
where $\varphi_{\alpha}(\textbf{y})$ is the atomic wave function in
the $\alpha$ state, $e$ is the electron charge absolute value,
$m_{p}$ and $m_{e}$ are the masses of the atomic core and electron,
respectively ($m=m_{p}+m_{e}$). In the case of the allowed dipole
transitions, in the linear order over $(\textbf{ky})\ll1$, one gets
\begin{equation}\label{eq.m-2.3}                                            %%%% ---- %%%% Eq. 3
    \sigma^{(1)}_{\alpha\beta}(\textbf{k})
    \approx i\textbf{k}\textbf{d}_{\alpha\beta},
    \quad\textbf{d}_{\alpha\beta}=e\int\textbf{y}
    d\textbf{y}\varphi_{\alpha}^{*}(\textbf{y})
    \varphi_{\beta}(\textbf{y}),
\end{equation}
where $\textbf{d}_{\alpha\beta}$ denotes the atomic dipole moment,
which is related to the transition $\alpha\leftrightarrow\beta$.

Considering that the density of the excited atoms is negligibly
small in comparison with the density of atoms in the ground state
(the case of a low intensity of the probe laser),  one can neglect
of the second term in Eq.~(\ref{eq.m-2.1}). Therefore, one gets
the expression
\begin{equation}\label{eq.m-2.4}                                           %%%% ---- %%%% Eqs. 4
    \epsilon^{-1}(\textbf{k},\omega;B)
    \approx 1+
    \sum\limits_{\alpha,\beta}
    {a_{\alpha\beta}}
    \left[\omega+\Delta\varepsilon_{\alpha\beta}(B)
    +i\gamma_{\alpha\beta}\right]^{-1}.
\end{equation}
Here the set of quantum numbers of ground and excited state are denoted by the
$\alpha$ and $\beta$ indexes, respectively.
As in this paper the limit of zero temperatures is considered,
the quantity $\gamma_{\alpha\beta}$ corresponds to the probability
of a spontaneous transition between the $\beta$ and $\alpha$ states.
Note that by the term ''ground state of an atom'' relates to all
sublevels of the ground state (for alkali metals these levels have
zero orbital number), which are splitted by the hyperfine and Zeeman
interaction. According to Eqs.~(\ref{eq.m-2.1}) and (\ref{eq.m-2.3}),
the quantity in Eq.~(\ref{eq.m-2.4})
\begin{equation}\label{eq.a11.5}                                           %%%% ---- %%%% new
    a_{\alpha\beta}=4\pi\nu_{\alpha}
    d_{\alpha\beta}^2/3
\end{equation}
defines the dependence of dispersion characteristics on the atomic
polarization and the density of atoms in the BEC state. Note also
that the dependence of dispersion characteristics on the intensity
of a static magnetic field~$B$ (see Eq.~(\ref{eq.m-2.4})) is mainly
defined by the energy interval $\Delta\varepsilon_{\alpha\beta}(B)$
between corresponding ground and excited states of an atom. Thus it
is necessary to study dependencies of the resonant
transitions on the field intensity in detail.

\section{Hyperfine structure of the sodium D$_2$ line in the external magnetic field}
Now it is important to obtain the energy spectrum of alkali-metal atoms in
the presence of a magnetic field with account of the hyperfine
interaction. To this end, one should consider the following Hamiltonian
(see also Ref.~\cite{Steck2000}):
\begin{equation}\label{eq.m-3.1}                                           %%%% ---- %%%% Eq. 05
    \hat{V}=A_{hfs}\hat{\textbf{I}}\cdot\hat{\textbf{J}}
    +B_{hfs}\hat{Q}+\mu_{B}(g_{I}\hat{I}_{z}
    +g_{J}\hat{J}_{z})B,
\end{equation}
where $A_{hfs}$ and $B_{hfs}$ are the hyperfine
constants related to the magnetic dipole and electric quadrupole
interaction of the nucleus and valence electron, respectively
($B_{hfs}=0$ for the states $J=1/2$). The quantities
$\hat{\textbf{I}}$ and $\hat{\textbf{J}}$ are the operators of the
nucleus and electron total angular momentum, respectively,
$\hat{I}_{z}$ and $\hat{J}_{z}$ are the operators of the projections
of these operators on the magnetic field $B$ direction, $\mu_{B}$ is
the Bohr magneton (in the units $\hbar=1$), and $g_{I}$ and $g_{J}$
are corresponding Land\'{e} factors. The operator $\hat{Q}$ can be
expressed in terms of the operators $\hat{\textbf{I}}$ and
$\hat{\textbf{J}}$ as follows:
\begin{equation}\label{eq.m-3.2}                                           %%%% ---- %%%% Eq. 06
    \hat{Q}=\frac{3(\hat{\textbf{I}}\cdot\hat{\textbf{J}})^2
    +\frac{3}{2}\hat{\textbf{I}}\cdot\hat{\textbf{J}}
    -I(I+1)J(J+1)}
    {2I(2I-1)2J(2J-1)}.
\end{equation}
Note that for the splitted levels of the ground state the spectrum
can be found analytically. At the same time, for the excited levels
(triple and more degenerated states in the total momentum projection
$m_F$, $\hat{\textbf{F}}=\hat{\textbf{I}}+\hat{\textbf{J}}$) it is
rather difficult to find corresponding analytical expressions. But
in this case one can find it numerically. One may verify that the
limiting cases of the numerical results lie in a good agreement with
the analytical formulas.

By the example of a sodium atom, it is possible to study the dependence of
$D_2$-line components on the magnetic field in detail. This line
corresponds to the transition between the 3$S_{1/2}$-state sublevels
and 3$P_{3/2}$-state sublevels. It is not difficult to obtain the
Zeeman splitting of these levels with account of the hyperfine
interaction (see Eqs.~(\ref{eq.m-3.1}) and (\ref{eq.m-3.2})) as
it is done in Ref.~\cite{Steck2000}.
As one can see, the resonant $D_2$ line
splits to a large number of components. Each component can be found
from the difference between corresponding excited and ground state
with the use of the selection rule for the quantum numbers~$F$ and
$m_F$ ($\Delta F=0,\pm1$; $\Delta m_F=0,\pm1$).

\begin{figure}
\includegraphics{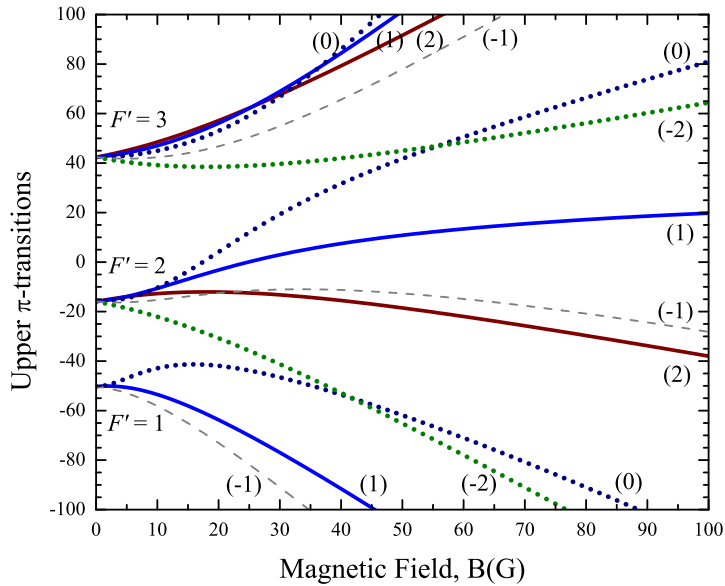}
    \caption{\label{fig.a11.01pitr}
    Components of the sodium $D_2$ line in the external static
    magnetic field ($F=2$; $\Delta F=0,\pm1$; $\Delta m_F=0$).
    The number in brackets corresponds to the momentum
    projection~$m_F$ of the quantum state.}
\end{figure}

Usually, depending on $\Delta m_F$ value, one can say about
\emph{linearly} or $\pi$\emph{-polarized} transitions ($\Delta m_F=0$),
and \emph{circularly} or $\sigma^{\pm}$\emph{-polarized} transitions,
$\Delta m_F=\pm1$ \cite{Scully2002}. Note that in the experiments
with ultraslow light the laser with the certain polarization is used.
Thus, one needs to consider the transitions between the states of an
atom corresponding to the certain polarization. Hence, in the next
calculations the response of the system to the
linearly-polarized pulse is considered.
For this radiation the resonance
frequencies correspond to the transitions $\Delta F=0,\pm1$; $\Delta
m_F=0$). The magnetic-field dependence for these components can be
found from the dependencies for the Zeeman splitting of the
ground and excited states (see, e.g.,
Ref.~\cite{Sly2009PLA}). Accounting the
mentioned selection rules, one can find the dependencies for the
''upper'' $\pi$-transitions (transitions from the ground state
with $F=2$) that are shown in Fig.~\ref{fig.a11.01pitr}.

\section{Dependence of the group velocity on the magnetic field intensity}
Now one can conclude that at certain intensities of magnetic fields
at the same frequency two different transitions are possible
(regions of the lines intersection, see Fig.~\ref{fig.a11.01pitr}).
These regions may be very interesting from the standpoint of the
experiments related to the ultraslow-light phenomenon. The reason is
that the value of the group velocity depends on the dispersion
characteristics of the medium. In particular, it depends on the
derivative of the refractive index by the frequency. Note that the
smaller the difference between two resonant frequencies, the
narrower the spacing between corresponding peaks of the refractive
index, and the steepness of the slope is larger. But this
characteristic does not mean that in the region of intersection one
should expect the maximal reduction of the group velocity. In this
case, the natural linewidth of the excited state component
(sublevel) plays an essential role.
These quantities for the excited sublevels can be obtained from the
relative hyperfine transition strength factors~$S_{FF'}$
\cite{Steck2000}, $\gamma_{FF'}=S_{FF'}\Gamma_{e}/2$, where for
sodium atom $\Gamma_{e}=9.795$~MHz. For example, if the atoms are prepared in
the ground states with $F=2$, one must take corresponding strength
factors for the ''upper'' transitions (see Fig.~\ref{fig.a11.01pitr}),
$S_{21}=1/20$, $S_{22}=1/4$, and $S_{23}=7/10$.

Next, analyzing dependencies that are shown in
Fig.~\ref{fig.a11.01pitr}, one can choose the frequency
of the linear-polarized light close to
the lowest intersection of the eigenfrequencies $\omega_1$
(dotted line in the lower part denoted by (0)) and
$\omega_2$ (dotted line in the lower part denoted by (-2)). In particular,
it is convenient to take $\omega^{(e)}=\omega_0+\Delta$,
where $\omega_0$ is $D_2$-line
resonance frequency without account of the hyperfine interaction
($\omega_0=508.848716$~THz), and $\Delta$ is the chosen detuning
($\Delta=-45.0$~MHz).
It is also necessary to consider that two corresponding ground-state
sublevels are occupied (quantum states $|1\rangle=|F=2,m_F=0\rangle$ and
$|2\rangle=|F=2,m_F=-2\rangle$).
In this case one can simplify calculations by
accounting only two resonant summands in (\ref{eq.m-2.4}). Really,
as one can conclude from Fig.~\ref{fig.a11.01pitr}, in this case the
major role play the transitions corresponding to the
eigenfrequencies $\omega_1$ and $\omega_2$. Therefore, one gets rather
simple expression for the permittivity,
\begin{equation}\label{eq.m-4.1}                                         %%%% ---- %%%% Eq. 7
    \epsilon^{-1}(\textbf{k},\omega ;B)
    \approx 1
    +\sum\limits_{j=1}^{2}
    {a_j}[\omega-\omega_j(B)+i\gamma_{j}]^{-1}.
\end{equation}
where the quantities $\gamma_{j}=S_{FF'}\Gamma_{e}/2$,
$a_j=4\pi\nu_j d_{FF'}^2/3$ (see Eq.~(\ref{eq.a11.5})), $d_{FF'}^2\approx
S_{FF'}(3.52er_0)^2$~\cite{Steck2000}, $r_0$ is the Bohr radius, and $e$ is the
elementary charge. The summand $j=1$ in this case corresponds
to the transition quantum numbers $F=2$, $F'=1$
and the summand $j=2$ corresponds to $F=2$, $F'=2$.

By introducing the real and imaginary parts of the permittivity
($\epsilon'$ and $\epsilon''$, respectively) it is easy to find the
relations defining the refractive index and damping factor
quantities:
$\epsilon'=n'^2-n''^2$ and $\epsilon''=2n'n''$.
In the transmission region the refractive index~$n'$ defines the
group velocity of the signal, and the damping factor~$n''$ defines
the intensity of the transmitted light,
\begin{eqnarray}\label{eq.m-4.3}                                         %%%% ---- %%%% Eq. 8
    v_{g}(\omega,B)=c
    \left\{n'(\omega,B)+\omega[\partial n'(\omega,B)/
    \partial \omega]\right\}^{-1},
    \\
    I(\omega,B)=I_{0}(\omega)\exp{[-n''(\omega,B)kL]},\label{eq.m-4.4}        %%%% ---- %%%% Eq. 9
\end{eqnarray}
where $I_{0}(\omega)$ is the spectral density
of the initial pulse, $L$ is the characteristic linear size of a condensed cloud.

\begin{figure}
\includegraphics{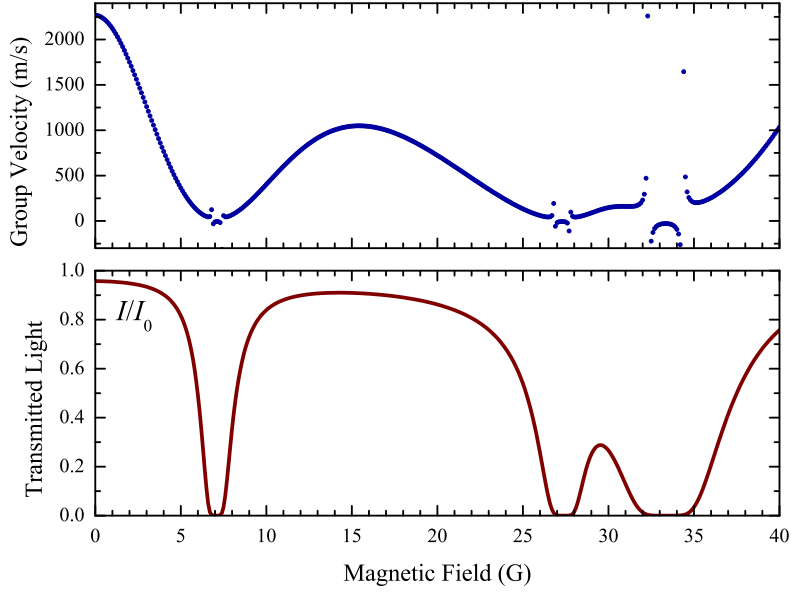}
    \caption{\label{fig.a11.02gvel}
    Dependencies of the group velocity and
    relative intensity on the external static magnetic
    field for the resonant light pulses with the constant
    detuning~$\Delta=-45.0$~MHz.}
\end{figure}

Next, by setting $\omega=\omega^{(e)}$,
$\nu_1=\nu_2=2\times10^{10}$~cm$^{-3}$, and $L=0.004$~cm, it is easy
to calculate the quantities (\ref{eq.m-4.3}) and (\ref{eq.m-4.4}) for the
different values of the intensity of a bias field. By the use of
these results one can get the dependencies that are shown in
Fig.~\ref{fig.a11.02gvel}.
As it is easy to see, these dependencies have a rather nontrivial
form. One can conclude that in some regions the velocity can be
reduced more than in ten times by the change of the field intensity
only in several gauss. Moreover, in some regions
($|\omega^{(e)}-\omega_i|<\gamma_i$) the group velocity has negative
values. But, as it is shown in the lower part of
Fig.~\ref{fig.a11.02gvel}, in these regions the signal is
mostly absorbed. The absorption rate can be made much lower by the
use of a gas with the lower density or with smaller characteristic
size~$L$. In this case one should observe the light pulses
propagating in a BEC with a negative group velocity. Note also that
the propagation of the light pulses with a negative group velocity
for similar systems is
well-studied both theoretically and experimentally (see
Ref.~\cite{Mac2003EPJ} for details).

\section{Possibility to filter the signals by the use
        of the ultraslow-light phenomenon in a BEC}

To show a possibility to filter optical pulses in a BEC in
the presence of the external static magnetic field it is
convenient to choose some model. In particular, when the
spectral density of the initial pulse has a normal
(Gaussian) distribution, it can be written as follows:
\begin{eqnarray}\label{eq.f-3.1}
    I_0(\omega)=\frac{I_0}{\sigma\sqrt{2\pi}}
    \exp\left[-(\omega-\omega_0)^2/2\sigma^2\right].
\end{eqnarray}
To get appreciable results, it is convenient
to choose $\omega_0=508.848052$~THz, $\sigma=50$~MHz
(the signal can effectively interact with several
hyperfine components of the sodium D$_2$ line, see
Fig.~\ref{fig.a11.01pitr}), and the linear polarization
of the signal.

\begin{figure}
\includegraphics{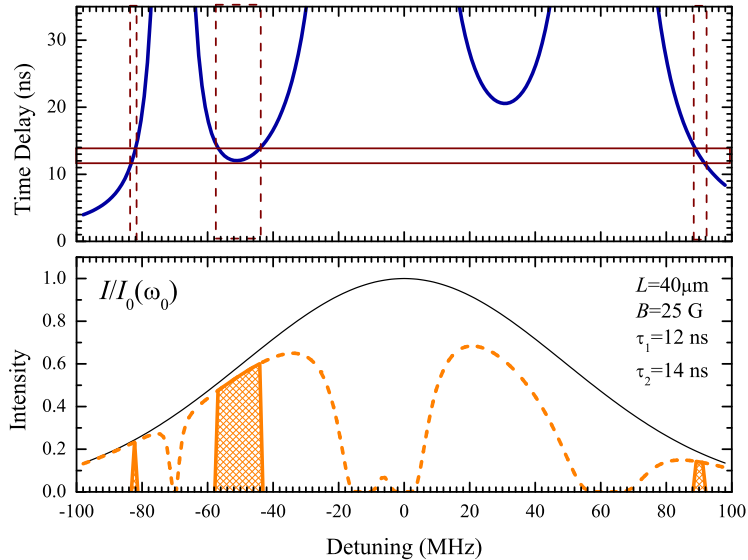}
    \caption{\label{fig.a11.03prop}
    Frequency dependencies of the time delay
    and relative intensity of the signal
    that propagates
    through a condensed gas with the occupied states
    $|F=2,m_F=1\rangle$ and $|F=2,m_F=2\rangle$.}
\end{figure}

Next, it is necessary to choose the gas properties.
For example, one can take the gas of sodium atoms
with two occupied ground states:
$|1\rangle=|F=2,m_F=1\rangle$ and
$|2\rangle=|F=2,m_F=2\rangle$. The resonant transitions
dependencies on the magnetic field
for these states are marked by solid bold lines in
Fig.~\ref{fig.a11.01pitr}. Next, taking for
the convenience $\nu_1=\nu_2=10^{12}$~cm$^{-3}$,
$B=25$~G, and $L=40~\mu$m,
in accordance with Eqs.~(\ref{eq.m-4.3}) and (\ref{eq.m-4.4}),
one can get the dependencies that are shown in
Fig.~\ref{fig.a11.03prop}.

Therefore, from Fig.~\ref{fig.a11.03prop} one can see
that different parts of the pulse have different time
delays resulting from the fact that the group velocity
of these parts in gas a with condensates has different values
due to the ultraslow-light phenomenon. It should be
noted that that at certain frequencies the signal
is mostly absorbed by medium. But this is not enough
(see dashed line in the lower part of
Fig.~\ref{fig.a11.03prop}) to
reach good parameters for filtering optical pulses.
One needs also to use additional techniques
for a registration of the signal only in the
certain time interval. As it easy to see from
Fig.~\ref{fig.a11.03prop} (bold solid line
in the lower part corresponds to the time interval
from 12 to 14~ns), by the use of the
mentioned method one can get filtered pulses
with relatively good parameters.

In conclusion, it is important to note that the cases,
which is studied in this paper, is demonstrative in some sense.
In the framework of the proposed approach one can obtain the
results for another polarizations, occupied states,
another frequency detunings, and also another types of alkali atoms.
Evidently, this fact expands the possibility for the experimental
observation of the effects that are studied in this work.

\ack
This work is partly supported by NAS of Ukraine, grant No. 55/51 -- 2009.
The author also want to thank Prof.~Dr.~Yurii~Slyusarenko for fruitful
discussions and Dr.~Gennadij~Sotnikov for a financial assistance
in visiting CEWQO~2009 conference.

\end{document}